\documentclass[iop,revtex4]{emulateapj}

\usepackage{graphicx}
\shortauthors{Jones \& West}
\shorttitle{M Dwarf UV activity}

\def\drnuvnum{186}
\def\drfuvnum{57}
\def\pmsunuvnum{371}
\def\pmsufuvnum{93}
\def\nuvnum{577}
\def\fuvnum{150}

\date{}

\begin{document}

\title{A Catalog of GALEX\altaffilmark{1} Ultraviolet Emission from Spectroscopically Confirmed M Dwarfs}
\author{David O. Jones\altaffilmark{2,3}, Andrew A. West \altaffilmark{4}}

\altaffiltext{1}{GALEX is operated for NASA by the California Institute of Technology under NASA contract NAS5-98034.}
\altaffiltext{2}{Corresponding author: djones@pha.jhu.edu}
\altaffiltext{3}{Department of Physics and Astronomy, The Johns Hopkins University, 3400 North Charles Street, Baltimore, MD 21218}
\altaffiltext{4}{Astronomy Department of Boston University, 725 Commonwealth Avenue, Boston, MA 02215, USA}

\begin{abstract} 

We present a catalog of GALEX Near-UV (NUV) and Far-UV (FUV) photometry for the Palomar/MSU and SDSS DR7 spectroscopic M dwarf catalogs.  The catalog contains NUV measurements matched to \nuvnum\ spectroscopically confirmed M dwarfs and FUV measurements matched to \fuvnum\ spectroscopically confirmed M dwarfs.  Using these data, we find that NUV and FUV luminosities strongly correlate with H$\alpha$ emission, a typical indicator of magnetic activity in M dwarfs.  We also examine the fraction of M dwarfs with varying degrees of strong line emission at NUV wavelengths.  Our results indicate that the frequency of M dwarf NUV emission peaks at intermediate spectral types, with at least $\sim$30\% of young M4-M5 dwarfs having some level of activity.  For mid-type M dwarfs, we show that NUV emission decreases with distance from the Galactic plane, a proxy for stellar age.  Our complete matched source catalog is available online.

\end{abstract}


\section{Introduction}

The physical processes underlying the magnetic dynamos of low-mass stars are of great importance to a comprehensive understanding of the Milky Way's stellar populations.  In addition to being of interest to the understanding of stellar interiors and the creation of stellar magnetic fields, stellar activity has ramifications for the transient population of the Galaxy and the potential habitability of attending planets.

Magnetic processes in low-mass dwarfs give rise to emission from the X-ray to the radio, but at many wavelengths the strength and frequency of this emission is poorly constrained, as is its dependence on spectral type.  Magnetic activity is best understood at optical wavelengths, where large spectroscopic studies such as the Sloan Digital Sky Survey \citep[SDSS]{Y00,A09} have enabled a statistical treatment of M dwarf activity, particularly by using H$\alpha$ emission as a diagnostic ($L_{\textrm{H}\alpha}$/$L_{\textrm{bol}}$;\citealp{W04,W08,W11}).  \citet{W08} showed that activity in the optical varies as a function of both spectral type and stellar age, with a peak in the active fraction occurring near a spectral type of M8 \citep{R95,H96,G00,W04}.

The manner in which these relationships extend to magnetic activity at X-ray, ultraviolet, and radio wavelengths is not as well-determined due to the small numbers of low-mass stars that have been observed at these wavelengths.  However, several recent studies have examined the relationships between H$\alpha$ and non-optical emission using data from all-sky surveys.  In the X-ray, \citet{R06} combined ROSAT data \citep{V99} with sources in the Two Micron All Sky Survey (2MASS; \citealp{S06}) to identify nearby M dwarfs for spectroscopic follow-up in order to examine the relationship between X-ray and H$\alpha$ luminosity.  They found that X-ray luminosity ``saturates'' (that there is an upper limit to the amount of quiescent X-ray emission) at a value of log($L_X$/$L_{\mathrm{bol}})\sim-3$.  \citet{S09} used emission from ROSAT as a signature of youth to build a census of low-mass stars within 25 pc.  They combined the ROSAT data with spectroscopic follow-up to identify spectroscopic binaries and estimate stellar ages, surface gravities, and magnetic activity.  In addition, \citet{C08b} used archival data from the Extended \textit{Chandra} Multiwavelength Project (ChaMP; \citealp{K04,K07}) to identify X-ray emitting stars and calculate a linear relationship between the H$\alpha$ and X-ray fractional luminosities in M dwarfs. \citet{C08b} also observed a decrease in the magnetic activity with age.  Because the work of \citet{C08b} was limited by having H$\alpha$ measurements for only $\sim$100 stars, further analysis is required to fully constrain the relationship between X-ray and H$\alpha$ emission in low-mass stars.  

At other wavelengths, \citet{Mclean12} used Very Large Array (radio wavelength) observations to derive a relation between radio emission and rotation for stars with spectral types M0-M6.  \citet{Harding13} found a correlation between radio emission and optical variability for late M and early L dwarfs, although the physical mechanism responsible for this phenomenon has yet to be determined.  In the ultraviolet regime, \citet{Wal08} used \textit{Hubble Space Telescope} (\textit{HST}) data to observe the near-ultraviolet (NUV; 1750-250\AA) emission of 33 M dwarfs.  Comparing NUV to optical and X-ray emission, they did not find a  universal trend between the optical and UV emission.   However, they did identify a clear correlation between \ion{Mg}{2} flux and ROSAT X-ray flux.  \citet{France13} also used \textit{HST} to image six M dwarf exoplanet hosts in the UV to characterize the radiation field incident on the planets.  They found substantial line variability on scales of $\sim$100-1000 seconds ($\sim$50-500\%), Ly$\alpha$ emission ($\sim$37-75\% of 1150-3100\AA\ emission), and hot H$_2$ gas of photospheric or possibly planetary origin.

These previous studies of M dwarf emission in the UV were limited by small sample sizes that restricted their statistical significance.  However, data from the Galaxy Evolution Explorer (GALEX; \citealp{M05,M07}) present a unique opportunity to constrain the ultraviolet properties of a much larger sample of low-mass stars, albeit using only broad photometric filters.  GALEX imaged $\sim$2/3 of the sky in its NUV and far-ultraviolet (FUV; 1350-1750 \AA) bands.  Although the primary purpose of GALEX was to examine the UV properties of extragalactic objects, several Galactic studies have used it to measure stellar UV emission.  \citet{W07} used GALEX data in order to examine the characteristics of M dwarf flares at UV wavelengths.  \citet{S11} and \citet{R11} used GALEX to show that ultraviolet activity is an indication of youth in low-mass stars and can be used to search for members of young moving groups in the solar neighborhood.  \citet{Shkolnik14} expanded on this analysis, showing that NUV activity as a function of age remains constant for stars younger than $\sim$300 Myr, similar to what has been observed for X-ray emission.  \citet{F11} explored the use of GALEX data as a stellar activity indicator for a calibration sample of nearby stars within 50 pc.  They presented preliminary relations between an optical Ca II activity indicator (R$'_{HK}$) and the GALEX UV flux and detected evidence for a subtle correlation between NUV flux and stellar age across a variety of stellar types.  \citet{F11} did not investigate other optical activity tracers such as H$\alpha$ or the fractions of UV-active stars.

Several recent studies have also examined the GALEX UV properties of M dwarfs within $\sim$50 pc of the sun.  \citet{Stelzer13} looked at the UV and X-ray properties of M dwarfs within 10 pc to find that UV chromospheric emission is connected to other activity indicators with a power-law dependence.  They find the range in activity for a given spectral type is $\sim$2-3 dex, which peaks in M4 dwarfs.  \citet{Ansdell15} used the \citet{Lepine11} magnitude-limited catalog of $J<10$ early-type M dwarfs and found a correlation between H$\alpha$ equivalent width and NUV$-$Ks magnitude, as well as an average activity level that remains constant until $\sim$200 Myr and then declines, a result that agrees with UV studies of young moving group members \citep{Shkolnik14}.  While \citet{Stelzer13} found that all M dwarfs in their 10 pc sample showed NUV emission above photospheric levels, \citet{Ansdell15} identified a population of M dwarfs showing only basal emission, the likely source of which is NUV line emission from the upper chromosphere present in many or all M dwarfs.   When examining UV activity, it is likely necessary to account for this emission.  We note that while \citet{Shkolnik14} and \citet{Ansdell15} use ``saturation'' to refer to the plateau of activity as a function of age for young stars, in this work we discuss saturation in the context of an upper limit to the level of emission in our most optically-active stars (e.g. \citealp{R06})

In this study, we take an extensive look at UV activity in both nearby and more distant M dwarfs.  We match broad band GALEX data to stars in the Palomar/MSU nearby star spectroscopic survey \citep{R95}, a local sample similar to that of \citet{Ansdell15}, and the SDSS Data Release 7 M dwarf spectroscopic sample (DR7; \citealp{W11}), which probes an older stellar population.  Using these catalogs, we were able to measure the UV properties of \nuvnum\ low-mass dwarfs across a wide range of distances and spectral types.  For the first time, we have examined the fraction of chromospherically-emitting M dwarfs as a function of spectral type and age, as well as the correlation between fractional NUV luminosity and fractional H$\alpha$ luminosity.  These diagnostics uncover a different set of physical relations than studies using R$'_{HK}$, H$\alpha$ equivalent width, and mean NUV luminosity as a function of age.

\begin{figure}
\plotone{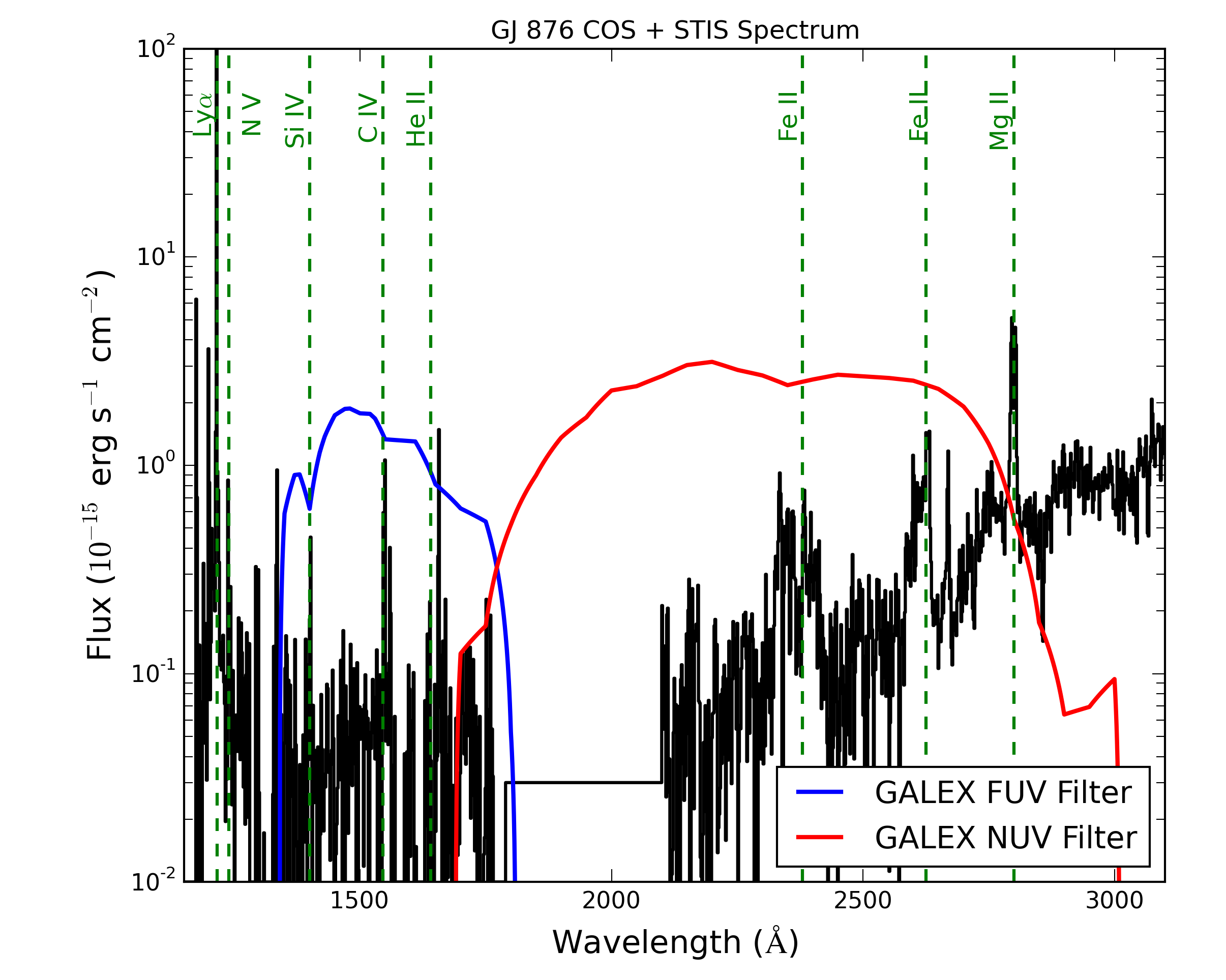}
\caption{The spectrum of Gl 876 from \citet{France13}, a moderately UV-active M4 dwarf imaged with the \textit{Hubble Space Telescope}'s Space Telescope Imaging Spectrograph (STIS) and Cosmic Origins Spectrograph (COS).  We have over-plotted the GALEX NUV (red) and FUV (blue) throughput curves (with an arbitrary scale factor).  We've marked some of the prominent emission line features visible in the GALEX NUV and FUV bands that trace M dwarf UV activity (green dashed lines).  The flux from 1760 $-$ 2100 \AA\ is absent as it is too faint to be detected by the STIS G230L grating.  For GJ 876, the GALEX NUV and FUV bands measure fluxes of 31.64$\pm$2.52 $\mu$Jy and 3.44$\pm$1.14 $\mu$Jy, respectively.  Given that GJ 876 is an M4 dwarf at a distance of 4.7 pc, we deduced approximate fractional NUV and FUV luminosities of log($L_{\mathrm{NUV}}$/$L_{\mathrm{bol}}$)$\sim$1.87$\times 10^{-5}$ and log($L_{\mathrm{FUV}}$/$L_{\mathrm{bol}}$)$\sim$1.59$\times 10^{-6}$.}
\label{fig:spec}
\end{figure}

In studying the UV emission with GALEX broadband filters, we trace a diverse group of spectral features.  Figure \ref{fig:spec} shows the moderately active, GALEX-detected, M dwarf spectrum of GJ 876 from \citet{France13} to demonstrate the UV line emission features (tracers of UV activity) that fall within the GALEX NUV and FUV passbands\footnote[1]{This work has made use of the MUSCLES M dwarf UV radiation field database.}.  GJ 876 has GALEX NUV and FUV fluxes of 31.64$\pm$2.52 $\mu$Jy and 3.44$\pm$1.14 $\mu$Jy, respectively.  In the FUV, the continuum emission is low, and line emission dominates the filter.  In the NUV, the continuum is present, along with \ion{Fe}{2} and \ion{Mg}{2} features.  Therefore, the GALEX NUV and FUV bands contain, and are often dominated by, tracers of stellar activity (emission lines), allowing us to expect a strong correlation between GALEX UV luminosities and magnetic activity tracers in the optical regime (\S4.1).  In addition, we can be confident that stars with high fractional UV luminosities must have strong line-emission present in the GALEX bands, allowing us to examine how the frequency of active stars depends on spectral type and distance from the Galactic plane (a tracer of stellar age; \S4.2).  GJ 876 also lacks H$\alpha$ emission or absorption, demonstrating that this moderate level of UV activity does not always correlate with optical line emission.

We describe the GALEX data along with the PMSU and SDSS DR7 M dwarf catalogs in \S2.  We describe the process of matching the GALEX data to the M dwarf samples and our analysis in \S3.  We present our results in \S4, and discuss our conclusions in \S5.

\section{Data}

The high quality photometric and spectroscopic data from SDSS over large solid angles have been useful for many large Galactic and extragalactic surveys.  For low-mass stars, the  SDSS has produced  photometric and spectroscopic samples of more than 30 million and 70,000 M dwarfs respectively \citep{B10,W11}.  Low-mass stars in SDSS have been used as tracers of Galactic dynamics \citep{B07a,B11,F09} Galactic extinction \citep{J11} and to investigate several aspects of the stars themselves, including but not limited to, the stellar mass and luminosity functions \citep{B10,C08}, properties of magnetic activity \citep{W04,W08}, and the low-metallicity population of the Galactic disk \citep{Savcheva14}.  

To compare magnetic activity in the optical with UV activity, we matched the SDSS Data Release 7 (DR7; \citealp{A09}) M dwarf spectroscopic catalog \citep{W11} to GALEX data from Data Releases 6 and 7.  The SDSS spectra were observed using twin fiber-fed spectrographs that collected 640 simultaneous observations.  Individual exposure times of $\sim$15-20 minutes were co-added for total exposure times of $\sim$45 minutes, producing medium resolution spectra with R $\sim$ 2000 \citep{Y00}.  The DR7 M dwarf catalog consists of 70,841 SDSS M dwarfs with spectral types verified by eye.  To measure optical stellar activity, the catalog includes (among other quantities) fractional H$\alpha$ luminosities ($L_{\textrm H\alpha}$/$L_{\textrm{bol}}$) for each star in the sample and the magnetic (H$\alpha$) activity state of each star (with ``active'' and ``inactive'' flags).  Distances were measured for all stars using the photometric parallax technique described in \citet{B10} and corrected for dust \citep{J11}, and have an accuracy of $\sim$18\%.  These uncertainties are distance-independent, as they result from photometric parallaxes and are dominated by variance in the M dwarf luminosity function.  See \citet{W11} for more details on the sample selection and the value added quantities measured for the SDSS DR7 M dwarf sample.

The SDSS sample contains a large number of M dwarfs.  However, the closest stars, for which we have the best chance of observing UV emission, often saturate SDSS detectors.  We therefore supplemented the SDSS sample with the Palomar/MSU Nearby-Star Spectroscopic Survey (PMSU; \citealp{R95}), which contains 1,684 nearby low-mass stars (1,415 M dwarfs) as part of the northern sample ($\delta>-30^{\circ}$) and 282 nearby low-mass stars (228 M dwarfs) as part of the southern sample.  These data contain spectral types, proper motions, magnitudes, distances, and spectra for most of the stars in the sample and provide a census of young, low-mass objects in the solar neighborhood.  The vast majority of stars in the sample lie within 50 pc with a median distance of $\sim$20 pc.  For active stars in the sample (dMe), photometry is given by \citet{H96}.  PMSU distances come from a variety of sources; many were measured by the Hipparchos Satellite, while others were derived using spectrophotometric parallax.  The trigonometric parallax-based distances in this sample have uncertainties $\lesssim$5\%, while the spectrophotometric parallax distances have uncertainties of up to 30\%.

We matched these two catalogs to the GALEX data, which consist mainly of two large-area surveys.  The All-Sky Imaging Survey (AIS) has surveyed $\sim$2/3 of the sky to a depth of NUV $m_{AB}\sim20.5$ and the Medium Imaging Survey (MIS) has imaged $\sim$1/10 of the sky to a depth of NUV $m_{AB}\sim23$.  The data contain images in a near-UV band from $\sim$1750-2750 \AA \ and a far-UV band from $\sim$1350-1750 \AA.  More information about the GALEX data releases can be found in \citet{M05} and \citet{M07} as well as at the GALEX website, part of the Mikulski Archive for Space Telescopes (MAST)\footnote[2]{http://galex.stsci.edu/.}.

\section{Analysis}

To match sources in the GALEX data with the SDSS DR7 M dwarf sample \citep{W11}, we queried the GALEX Data Release 6/7\footnote[3]{The GALEX GR6/7 is available at http://mastweb.stsci.edu/gcasjobs/.} to identify UV sources within 5 arcsec of the stellar positions.  We found matches to $\sim$3,511 SDSS sources.  Some SDSS stars had multiple GALEX counterparts; to eliminate chance alignments and remove multiple detections, we used the proper motions in the DR7 M dwarf sample \citep{M04} to determine the position for each star on a given observation date.  Using these positions, we were able to eliminate most of the multiple matches, keeping only the UV source that was closest to the position of the DR7 M dwarf at the date that GALEX observed it.  We removed all GALEX matches with the nuv\_artifact flag (for NUV analysis) or fuv\_artifact flag (for FUV analysis) greater than 1 and stars located more than 0.55 degrees from the center of the field of view, which are near the edge of the detector and have less reliable photometry \citep{M07}.  Only DR7 stars that reported good photometry (the ``GOODPHOT'' flag set to 1; see \citealp{W11}) and that had colors inconsistent with possible M dwarf - White dwarf binaries, were used.  Stars that reported FUV flux but no NUV detection were removed, due to the fact that M dwarfs are extremely unlikely to have FUV emission without having corresponding emission in the NUV (for example, see observed NUV and FUV fluxes in \citealp{S11}).  We also used the positional errors reported by GALEX to remove GALEX sources located more than 2 arcsec away from the corresponding SDSS M dwarf position to reduce the likelihood of spurious matches.



To analyze the activity of M dwarf - White dwarf binaries, we also investigated whether any of the 495 possible pairs identified in the full DR7 M dwarf sample could be found in GALEX.   Possible M dwarf - White dwarf binaries were selected from among the visually identified M dwarfs in the DR7 M Dwarf sample using the color cuts of \citet{S04}.  These binaries were then verified using the method of \citet{Morgan12}, which developed color cuts from GALEX, SDSS, UKIDSS \citep{Lawrence07}, and 2MASS \citep{Skrutskie06} photometry to reliably identify binaries.  Approximately 1/5, or $\sim$100, of the possible WD-dM pairs could be visually verified as binaries and 75 of these positions had been observed by GALEX.  Out of the 75 positions, we found matches for 42 of these pairs in GALEX after sample cuts, a \textit{much} higher detected fraction (56\%) than the fraction of SDSS M dwarfs detected overall.  Increased activity in WD-dM pairs has already been observed in the optical by \citet{Morgan12}, who proposed that such an increase in activity is a result of faster stellar rotation arising due to possible tidal effects, angular momentum exchange, or disk disruption.

To match GALEX to the PMSU sample, we first corrected the PMSU positions and proper motions using the SIMBAD Astronomical Database\footnote[4]{http://simbad.u-strasbg.fr/simbad/} and the \citet{Lepine11} M dwarf catalog.  SIMBAD's astrometry originates from a variety of sources, typically 2MASS \citep{S06} or Hipparcos (e.g. \citealp{Hog00}), and \citet{Lepine11} proper motions and astrometry originate from the SUPERBLINK survey \citep{Lepine02}.  Both are much more accurate than the original PMSU catalog.  We queried the GALEX DR6/7 CasJobs database to identify UV sources within 5 arcsec of the position of each PMSU star in each year of GALEX observations (2003-2013).  We kept only the closest GALEX match to a given star's positions during the years those positions were observed by GALEX.  We removed GALEX matches with the nuv\_artifact flag (for NUV analysis) or fuv\_artifact flag (for FUV analysis) greater than 1 and stars located more than 0.55 deg from the center of the field of view.  Although most PMSU positions and proper motions are accurate, $\sim$20\% of our sample had somewhat uncertain proper motions.  In spite of this, the positional uncertainties are typically accurate, allowing us to apply the same 2 arcsec angular separation cut that we used for SDSS M dwarfs.

Using the DR7 and PMSU spectra, we derived several stellar properties that were useful to quantify the UV activity.  First, we used a second-order polynomial relation to determine the absolute bolometric magnitude of our M dwarfs as a function of spectral type, which we then converted to stellar luminosity.  This relation was created by fitting to the 8-parsec stellar sample of \citet{RG97}.  Although most of the stars in this sample are inactive, and active stars may have higher bolometric luminosities, our derived values are $\sim$35\% lower than those found for the inactive early-type dwarfs in \citet{Mann13} and agree well with the values of \citet{Dieterich14} for late-type dwarfs.  We next calculated the H$\alpha$ luminosity of the DR7 and PMSU samples following \citet{W11}, which integrates over a line region of width 8\AA, and subtracts a the nearby continuum (see Table 1 and \S2.2 in \citealp{W11}).

We used the stellar distances and GALEX bandpass data to convert the GALEX NUV and FUV fluxes, measured in $\mu$Jy, to luminosities in erg s$^{-1}$.  Once the H$\alpha$, NUV, and FUV luminosities were measured, we divided these quantities by each star's bolometric luminosity in order to find the fractional luminosities emitted by each of these three components.

\subsection{Removing Contaminating Sources}
\begin{deluxetable*}{lcccc}
  \tabletypesize{\small}
  \tablecolumns{5}
  \tablewidth{0pt}
  \tablecaption{M dwarf samples after selection cuts.
    \label{table:samplecuts}}
  \tablehead{ & \multicolumn{4}{c} { Number of GALEX-matched M dwarfs}\\
                     & {PMSU NUV} & {PMSU FUV} & {SDSS NUV} & {SDSS FUV}
  }
  \startdata
  Total Unique Matches&877&265&3511&968\\*[2pt]
  \\[-9pt]
  $-$ FOV radius $>$ 0.55&91&16&501& 108\\*[2pt]
  $-$ artifact flag $>$ 1&159&28&451&191\\*[2pt]
  $-$ FUV without NUV&\nodata&11&\nodata&205\\*[2pt]
  $-$ Suspect SDSS photometry&\nodata&\nodata&207&32\\*[2pt]
  $-$ MD-WD binary&\nodata&\nodata&146&74\\*[2pt]
  $-$ $>$2 arcsec from survey position&99&26&1525&198\\*[2pt]
  $-$ $<$3$\sigma$ detection&80&56&339&37\\*[2pt]
  $-$ Binary/nearby source in the literature&77&35&\nodata&\nodata\\*[2pt]
  $-$ non-M dwarf source within 5.5 arcsec&0&0&156&66\\*[2pt]
  \hline\\[-6pt]
  Final Sample &371&93&186&57\\*[2pt]
  \enddata
\end{deluxetable*}

There are a variety of possible contaminants of our M dwarf sample.  The most likely source is contaminating background sources, as well as blended sources due to the $\sim$5.5 arcsec GALEX PSF.  Another possible contaminants is unresolved binaries (such as M dwarf - White dwarf binaries).

In the SDSS sample, most close binary sources have been discovered and removed from the sample via visual spectral classification.  Additionally, possible M dwarf - White dwarf binaries were removed via the method of \citet{Morgan12}, as discussed above.  In PMSU, we flagged all stars with known nearby sources using Digitized Sky Survey Images in conjunction with SExtractor \citep{Bertin96} to identify nearby companions in the images and remove those stars from our sample.  Second, we queried SIMBAD for known binary companions in the literature and removed these from our sample.  In total, only 4 DSS sources had companions within 5.5 arcsec, but 57 of our previously-identified M dwarf matches had companions in the literature.

We then estimated the probability of false matches (contaminating background sources) for both samples by querying GALEX for a series of false object positions within a few arcminutes of the true positions such that our false positions would likely have been observed with the same exposure times as the true positions.  For both surveys, we found a $\sim$0.5\% chance of detecting a source for a given random position.  In PMSU, this is only a minor concern; because over 25\% of PMSU stars are detected in GALEX, the false positive fraction is $\lesssim$2\% for the sample.  For active (dMe) stars, the detection rate is higher, and the false positive fraction is correspondingly lower.

For SDSS, contaminating background sources are much more of a concern.  Because we initially detected $\sim$1,000 sources that passed all cuts out of our 70,000 initial positions, up to 30\% of our matched sample could be false positives.  Although we were unable to completely alleviate this concern, we were helped by the fact that the vast majority of contaminating background sources should be detected in SDSS.  We therefore removed all stars from our sample that had SDSS-detected sources within 5.5 arcsec ($\sim$1 PSF FWHM).  Our random source list detected 376 random NUV matches within 1000 pc that passed all our cuts and when we removed all known SDSS sources, just 170 matches remained.  Finally, when we limited our results to 3$\sigma$ detections to remove noise fluctuation (we removed $<$3$\sigma$ detections from PMSU matches as well), 26 were left (compared to 186 stars in the full sample).  We found 146 spurious FUV matches, and all were removed when we removed all known SDSS sources.  For the NUV, our estimated contamination is 6\% for H$\alpha$-active stars and 14\% for H$\alpha$ inactive stars.

After performing these steps, we believe that the correlations discussed in this paper are robust, but note that the measured fraction of emitting stars is somewhat biased towards non-detections due to the 3$\sigma$ flux threshold and strict 2 arcsec matching radius, which is within the 2$\sigma$ positional uncertainty for some GALEX sources.  Table \ref{table:samplecuts} lists each cut we made on our sample, and the number of sources each step removed.  Our final catalog had \drnuvnum\ NUV-matched DR7 sources, \drfuvnum\ FUV-matched DR7 sources, \pmsunuvnum\ NUV-matched PMSU sources, and \pmsufuvnum\ FUV-matched PMSU sources.

\subsection{Examining the Changing Fractions of UV-Emitting M Dwarfs}
\label{section:3.1}
We examined the fractions of UV-emitting M dwarfs by determining the fraction of stars in our sample that showed NUV or FUV emission above specific threshold fractional luminosities.
Any UV emission greater than the expected continuum emission should be due to line emission, and therefore magnetic activity.  In \S4, we examine three different thresholds of UV emission that span the range of line emission levels for which there are non-zero fractions of stars both above and below the threshold, and which are not so faint that the GALEX emission is likely coming from the UV continuum.  The choice of which thresholds of UV emission to use is necessarily semi-arbitrary, as broad band photometry cannot precisely detect the presence or absence of emission lines, particularly for the stars in our sample with large distance uncertainties. 

\begin{figure}
\plotone{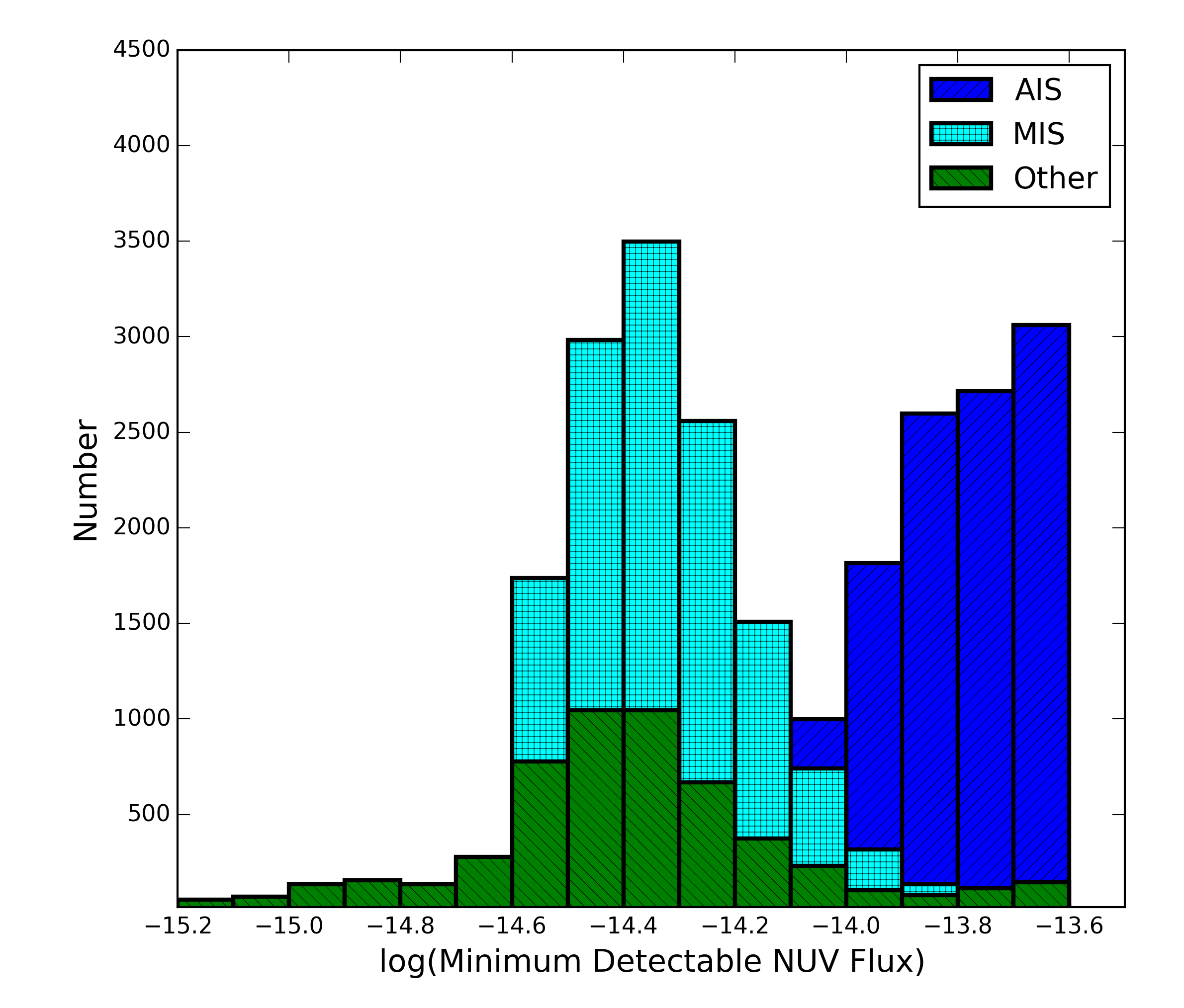}
\caption{The distribution of minimum NUV fluxes in log(erg s$^{-1}$ cm$^{-2}$) detectable by GALEX for all of the positions in the DR7 M dwarf sample.  The two peaks correspond to the different depths of the All-Sky Imaging Survey (AIS; NUV limiting magnitude $m_{AB}\sim20.5$) and the Medium Imaging Survey (MIS; NUV limiting magnitude $m_{AB}\sim23$).  The Deep Imaging Survey (DIS) and guest observer programs make up the broader green histogram.  Histograms of the DR7 sample in the FUV and of the PMSU sample in both the NUV and FUV show a similar bimodal appearance.}
\label{fig:min flux hist}
\end{figure}

For a statistical understanding of the role that magnetic activity plays in M dwarfs, the quiescent, undetectable stars are as important as the active ones. To determine the emitting fractions of M dwarfs, we first assessed the sensitivities of the different GALEX data products, which have different exposure times and magnitude limits.  We made certain that a given star was not detected merely because the exposure time was not long enough.  To measure whether a star with a certain UV luminosity could be detected, we identified the minimum observable UV fluxes of each field in which a M dwarf was located.  We first queried CasJobs to determine whether the positions had been observed by GALEX.  Out of 70,841 DR7 stars, only 8,022 had positions that had not been observed by GALEX, while 20\% ($\sim$395 stars) of the PMSU sample had not been observed.  For the positions that had been observed (detections and non-detections), we estimated the detection threshold as the flux uncertainty of the faintest neighboring source within 2 arcminutes.  Figure \ref{fig:min flux hist} shows the distribution of the minimum detectable NUV fluxes for the DR7 sample, which has two notable peaks.  The larger minimum fluxes are generally sources observed in the All-Sky Imaging Survey (AIS), while the smaller minimum fluxes are generally observed with the Medium Imaging Survey (MIS) or another smaller-area GALEX survey. 

\begin{figure*}
\plotone{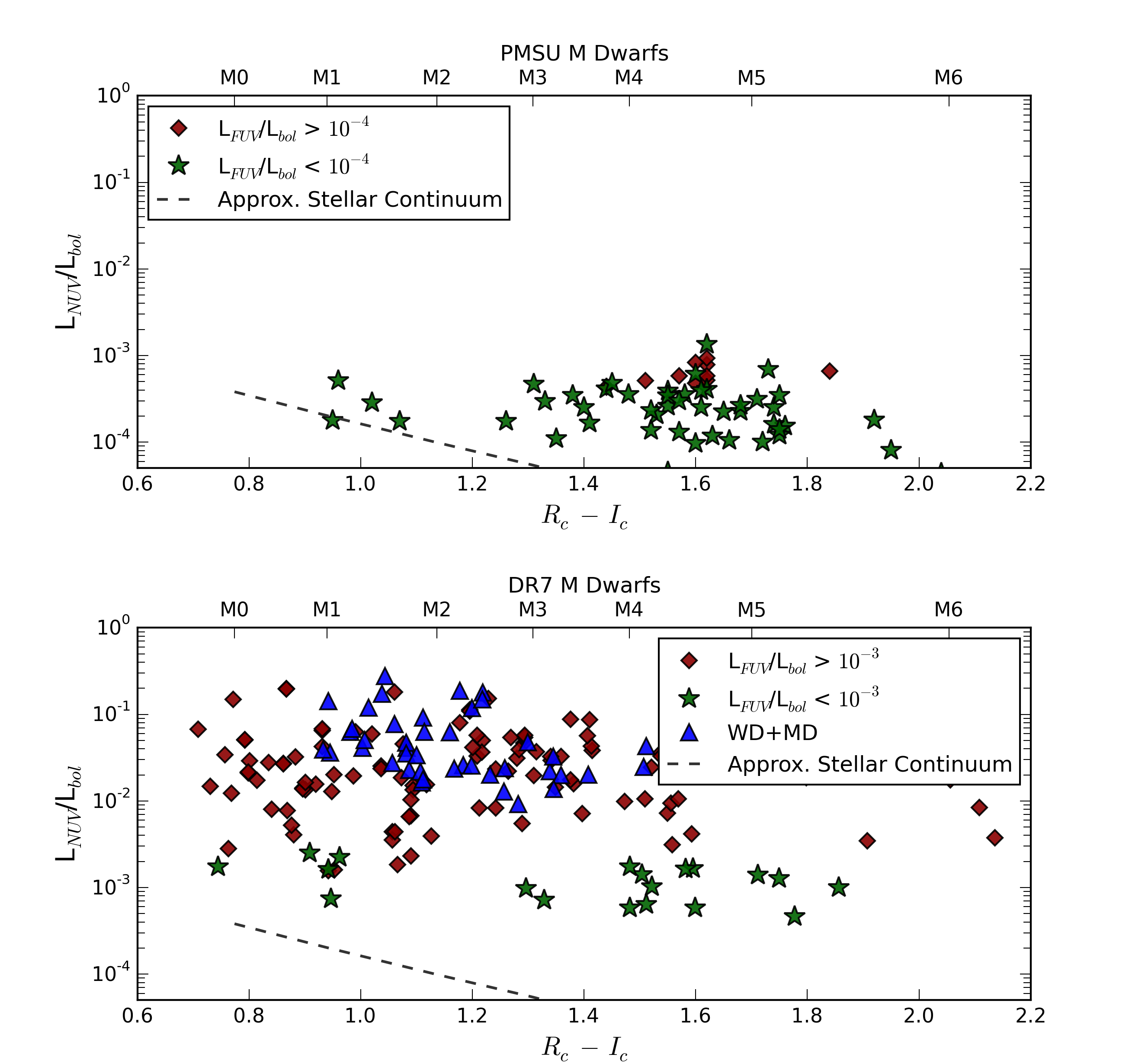}
\caption{$L_{\mathrm{NUV}}/L_{\mathrm{bol}}$ as a function of $R_c - I_c$ color for the H$\alpha$-active DR7 (top) and dMe PMSU (bottom) M dwarfs detected in both the NUV and FUV with GALEX.  We labeled the median $R_c-I_c$ color for various spectral types along the upper axes.  The DR7 sample is separated into stars with $L_{\mathrm{FUV}}/L_{\mathrm{bol}} > 10^{-3}$ (red diamonds), stars with $L_{\mathrm{FUV}}/L_{\mathrm{bol}} < 10^{-3}$ (green stars), and likely M dwarf - White dwarf binaries (blue triangles).  The PMSU sample is divided into stars with $L_{\mathrm{FUV}}/L_{\mathrm{bol}} > 10^{-4}$ (red diamonds), stars with $L_{\mathrm{FUV}}/L_{\mathrm{bol}} < 10^{-4}$ (green stars).  As expected, there is a pronounced separation in NUV luminosity between stars with low FUV activity and stars with high FUV activity.  M dwarf - White dwarf binaries appear much more UV emission, on average, than other stars.  We've indicated the approximate M dwarf NUV blackbody continuum luminosities for these colors based on the M dwarf temperatures given in \citet{R05} and the average colors of M0-M9 dwarfs in the DR7 and PMSU samples.  Some stars could have only continuum emission, but this is impossible to verify with these data due to stellar distance and temperature uncertainties.}
\label{fig:evplot}
\end{figure*}

For some stars, the NUV stellar continua should be detectable.  An M0 dwarf at 10 pc, for example, will have a GALEX NUV continuum flux of approximately $4 \times 10^{16}$ erg s$^{-1}$ based on an average temperature of 3,800 K \citep{R05}, which could be detected in some of our deepest fields.  The PMSU sample has 57 M0-M2 dwarfs within 10 pc, so we expect a handful of GALEX-detected dwarfs that don't have significant NUV activity.  When examining the fraction of M dwarfs with line emission, using broadband photometry alone, we must be sure to avoid stars such as these.

The sensitivity of GALEX allows us to examine the fraction of M dwarfs with $L_{\mathrm{NUV}}/L_{\mathrm{bol}} > 10^{-4}$, $>5\times10^{-4}$, and $>10^{-3}$.  Higher NUV luminosity limits would eliminate the majority of M dwarfs from the sample, and lower luminosity limits begin to observe the M dwarf NUV continuum.  We used the GALEX detection thresholds in each field to determine which M dwarfs were detectable at each UV emission level in conjunction with a Monte Carlo sampling of the NUV fractional luminosity uncertainties, which contribute to uncertainty in both the emission and the emitting threshold of a given star.  We subtracted the \citet{Ansdell15} basal NUV emission from our detections and minimum thresholds.

In addition to these criteria, positions that were 0.55 degrees or more from the center of the GALEX field of view or positions for which the closest match reported an nuv\_artifact flag (for NUV analysis) or fuv\_artifact flag (for FUV analysis) greater than one were not used, in agreement with the criteria we set above for reliable GALEX detections.  We also excluded positions whose nearest neighbor was $>$ 0.55 degrees or more from the center of the GALEX field of view, due to the fact that the reported photometry of these neighbors could be erroneous.  This slightly biases the measured activity fractions (tending to raise them), due to the fact that this quality cut was applied exclusively to non-detections, but because essentially every position within 0.55 degrees of field center had a neighbor within 0.55 degrees of the field center ($\sim$99\%), the effect is likely negligible.  In our matched source catalog, we include GALEX measurements for stars that don't meet any of the fractional luminosity thresholds discussed above.

\section{Results}

\begin{figure}
\epsscale{1.2}
\plotone{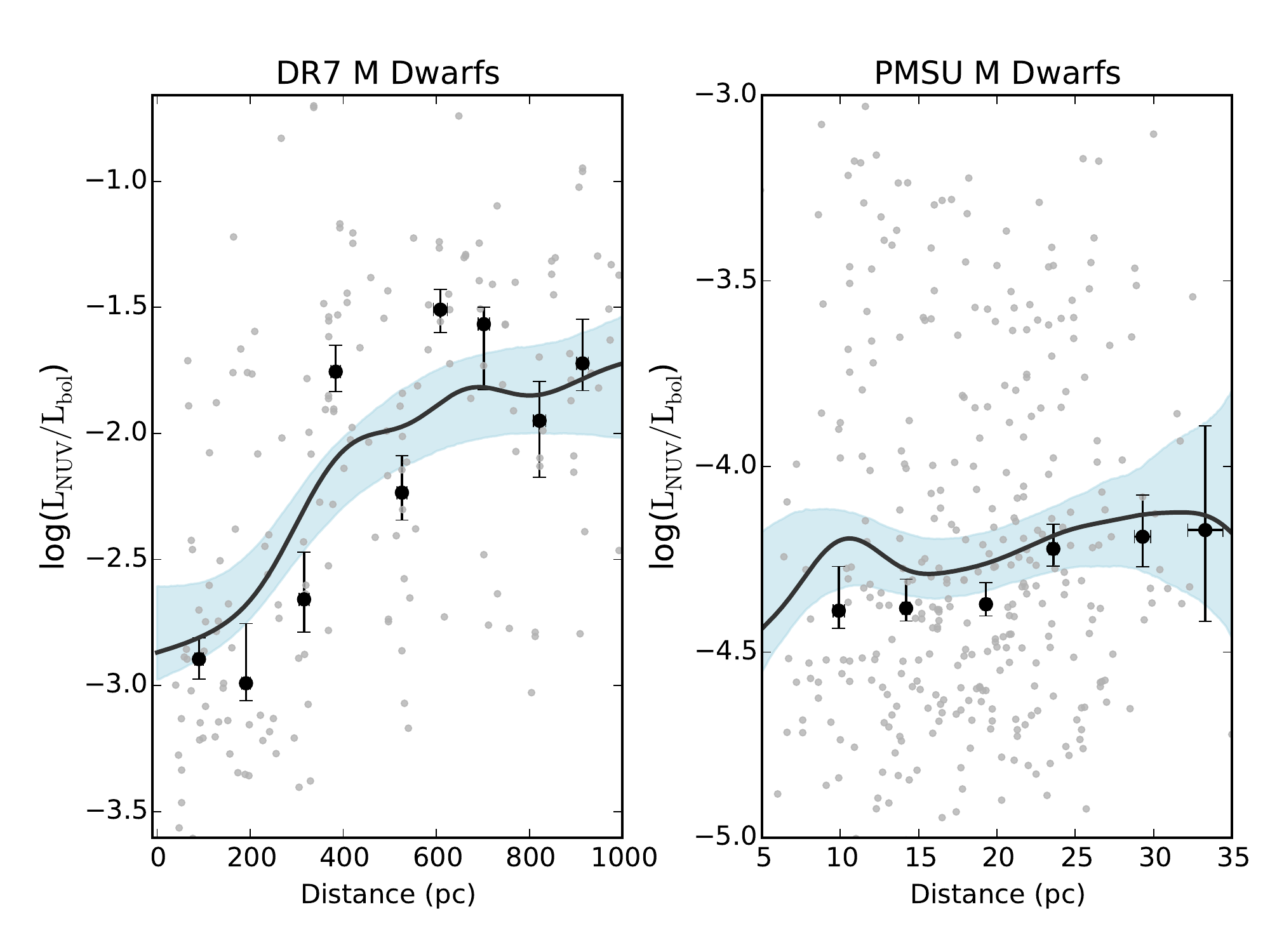}
\caption{The fractional NUV luminosity, log($L_{\mathrm{NUV}}/L_{\mathrm{bol}}$), of DR7 (left) and PMSU (right) M dwarfs detected in GALEX as a function of distance.  We show both individual stars and median bins.  A selection effect is likely responsible for the upward trend in the DR7 data.  PMSU M dwarfs appear to have similar NUV luminosities at a variety of distances, signifying that there is a reduced or negligible selection effect for nearby stars.  Blue curves show bootstrapped 95\% confidence intervals from non-parametric spatial averaging with a black line indicating the mean.}
\label{fig:nuvdist}
\end{figure}

Our final matched sample consists of \drnuvnum\ NUV-matched and \drfuvnum\ FUV-matched DR7 M dwarfs and \pmsunuvnum\ NUV-matched and \pmsufuvnum\ FUV-matched PMSU M dwarfs.  Tables \ref{table:DR7 data} and \ref{table:PMSU data} contain the first ten rows of our matched DR7 and PMSU samples, respectively, and include object IDs, positions, spectral types, and NUV, FUV and H$\alpha$ fractional luminosities for the M dwarfs in our samples.  The full catalogs can be found online\footnote[5]{\url{http://www.pha.jhu.edu/~djones/GALEX.html}}.

Figure \ref{fig:evplot} shows H$\alpha$-emitting, GALEX-detected M dwarfs in our matched source catalog, with $L_{\mathrm{NUV}}/L_{\mathrm{bol}}$ as a function of $R_c - I_c$ color for the DR7 M dwarfs (top) and the PMSU sample (bottom).  For the SDSS stars, we converted $r - i$ to $R - I$ color using the empirical relations of \citet{Jordi06}.  We computed the median $R-I$ color for each spectral type and labeled these along the top axis.  Note that spectral types M7-M9 are redder than the limit of the x axis; only a handful of active stars with spectral types of M7 and later were matched to GALEX due to the lower luminosities of these stars in SDSS and the small number of late-type M dwarfs contained in PMSU.  We distinguished between stars with high FUV activity (red diamonds) and low FUV activity (green stars).  As expected, stars with greater FUV activity had significantly greater NUV activity (NUV flux/FUV flux correlation coefficients for the PMSU and DR7 samples are 0.977 and 0.998, respectively).  Stars with undetected FUV activity (omitted for visual clarity) appear to be spread throughout the sample, likely due to the greater difficulty of M dwarf FUV detections for more distant stars and the 2009 malfunction of the FUV detector.  We also found that the stars in the DR7 sample flagged as M dwarf - White dwarf binaries (from \citealp{Morgan12}; blue triangles) tended to be significantly more NUV-luminous on average than other stars.  \citet[their Figure 3]{S11} find similar results for stars in the NStars 25 pc sample \citep{R07} detected in GALEX.  The dashed line in Figure \ref{fig:evplot} shows the approximate NUV blackbody continuum emission for these stars as a function of color, based on M dwarf temperatures from \citet{R05} and the typical colors for M dwarfs in the DR7 and PMSU samples.  Some stars appear to emit less than the approximate NUV continuum, but this is likely due to our uncertainties.  For these stars, we may have detected only the UV continuum, but stellar distance and temperature uncertainties make this impossible to determine with these data.

We caution that distance has a significant effect on our sample.  Because stars at greater distances become increasingly difficult to detect, we found that the NUV luminosities became increasingly larger as the distances increased.  This is opposite of the expected effect for an unbiased sample, which would tend to probe older populations and potentially less magnetically active stars at greater distances, as these stars would tend to lie farther from the Galactic plane (because most SDSS sight-lines are near the north Galactic cap).

Figure \ref{fig:nuvdist} shows the dependence of the observed $L_{\mathrm{NUV}}/L_{\mathrm{bol}}$ for both DR7 and PMSU M dwarfs as a function of distance.  In this and future figures we used the median $L_{\mathrm{NUV}}/L_{\mathrm{bol}}$ values for the stars as the central value in each distance bin and we measured errors using the dispersion around the median value divided by the square root of the number of stars in each bin.  We also show a non-parametric regression using spatial averaging, with 95\% confidence intervals from bootstrap resampling.  We found that both the NUV and FUV fractional luminosities for the DR7 sample increased with distance, indicating the presence of a selection effect (intrinsically brighter objects are easier to see at large distances).  This indicated that most of the DR7 stars detected in GALEX needed to not be only active, but highly active to be detected.  However, PMSU stars show no significant trend with distance due to the small range of distances probed by the sample.  Regardless, we were able to examine the relationship between UV and H$\alpha$ fractional luminosities, with the caveat that the DR7 data (and the PMSU data to a lesser extent) could only be used to probe the highly UV-luminous part of the relation.  Throughout this section, we have verified that our results are consistent using a variety of distance limits to ensure that our results were not affected by spurious matches beyond several hundred parsec (we nominally restrict our sample to $<$1000 pc).

\subsection{Correlating Optical and UV Activity}

\begin{figure}
\epsscale{1.2}
\plotone{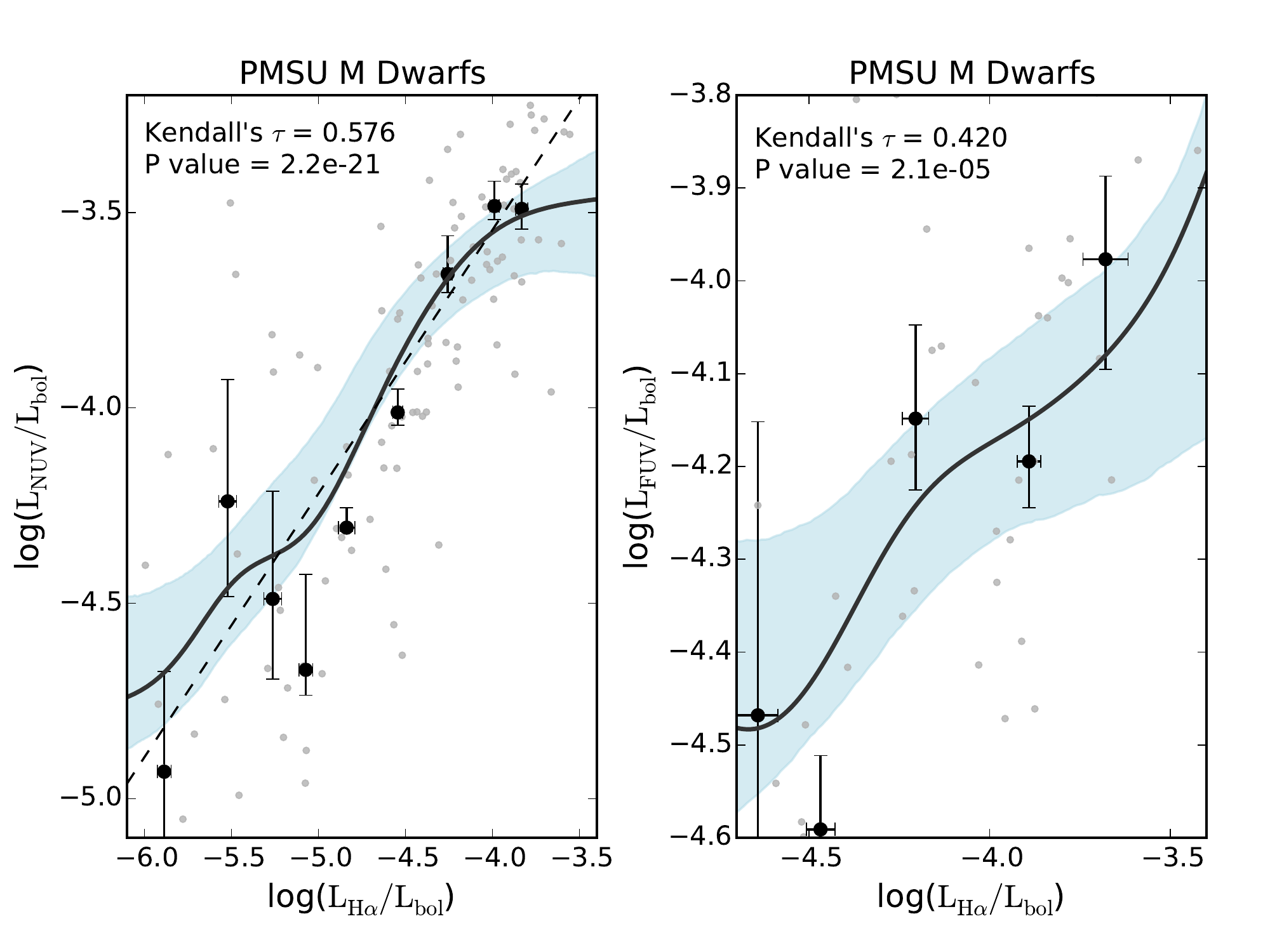}
\caption{The fractional NUV luminosity (left; log($L_{\mathrm{NUV}}/L_{\mathrm{bol}}$)) and FUV luminosity (right; log($L_{\mathrm{FUV}}/L_{\mathrm{bol}}$)) plotted against the fractional H$\alpha$ luminosity, Log($L_{\mathrm{H\alpha}}/L_{\mathrm{bol}}$), for stars in the PMSU sample (individual stars and median bins).  There is a clear trend on the left, indicating that NUV emission is a good indicator of overall stellar activity. A trend in the FUV also exists (2.2$\sigma$ significance), although less FUV data exist for the PMSU sample.  We show the best-fit lines to the data, as well as a non-parametric spatial averaging with 95\% confidence intervals from bootstrapping (blue curves, with the mean in black).  The best-fit dashed line to the NUV-H$\alpha$ relation is given by equation 1.}
\label{fig:pmsu uvhalpha}
\end{figure}

Aside from creating a catalog of M dwarfs with UV emission, one of the principal goals of this study was to better understand how optical activity, traced by H$\alpha$, correlates with NUV and FUV activity.  To investigate this correlation, we compared the NUV and FUV emission with the H$\alpha$ flux in stars with H$\alpha$ emission.  \citet{Ansdell15} discovered a strong correlation between NUV luminosity and H$\alpha$ equivalent width for M0-M3 dwarfs.  Our results include later spectral types but are qualitatively similar, with the caveat that we use H$\alpha$ fractional luminosity to probe this relation, a quantity less dependent on spectral type or color than H$\alpha$ equivalent width (used in \citealp{Ansdell15}).  We subtracted the expected basal chromospheric emission following \citet{Ansdell15} to remove the approximate contribution of chromospheric emission lines to the UV flux.  For most stars in our sample, this basal emission is well below the detection threshold.

Figure \ref{fig:pmsu uvhalpha} shows a strong correlation between NUV and FUV luminosities and the H$\alpha$ luminosities of the PMSU sample.  However, we do not see a correlation with DR7 stars, likely because the close proximity of PMSU stars enable intrinsically fainter UV detections; the minimum detected UV flux is 1-2 orders of magnitude lower for the PMSU sample than for the DR7 stars, demonstrating that when we examine SDSS data, only the most active stars can be detected.  The PMSU sample shows a strong correlation between NUV and H$\alpha$ (Figure \ref{fig:pmsu uvhalpha}; left panel) that is independent of spectral type (although the PMSU sample is largely composed of spectral types M3 and M4).  This correlation is expected, as it is thought that both UV and H$\alpha$ emission are good indicators of overall stellar activity.  The best fit line of the PMSU NUV to H$\alpha$ luminosity relation is:

\begin{equation}
\textrm{log}(L_{\mathrm{NUV}}/L_{\mathrm{bol}}) = 0.67 \times \textrm{log}(L_{\mathrm{H\alpha}}/L_{\mathrm{bol}})-0.85,
\end{equation}

\noindent where the formal 1-$\sigma$ uncertainties in the slope and Y-intercept are 0.12 and 0.57, respectively.

The right panel in Figure \ref{fig:pmsu uvhalpha} shows the correlation between PMSU FUV and H$\alpha$ luminosities, which is less strong than the NUV-H$\alpha$ correlation but still indicates a trend (2.2$\sigma$ significance).

\begin{figure}
\epsscale{1.2}
\plotone{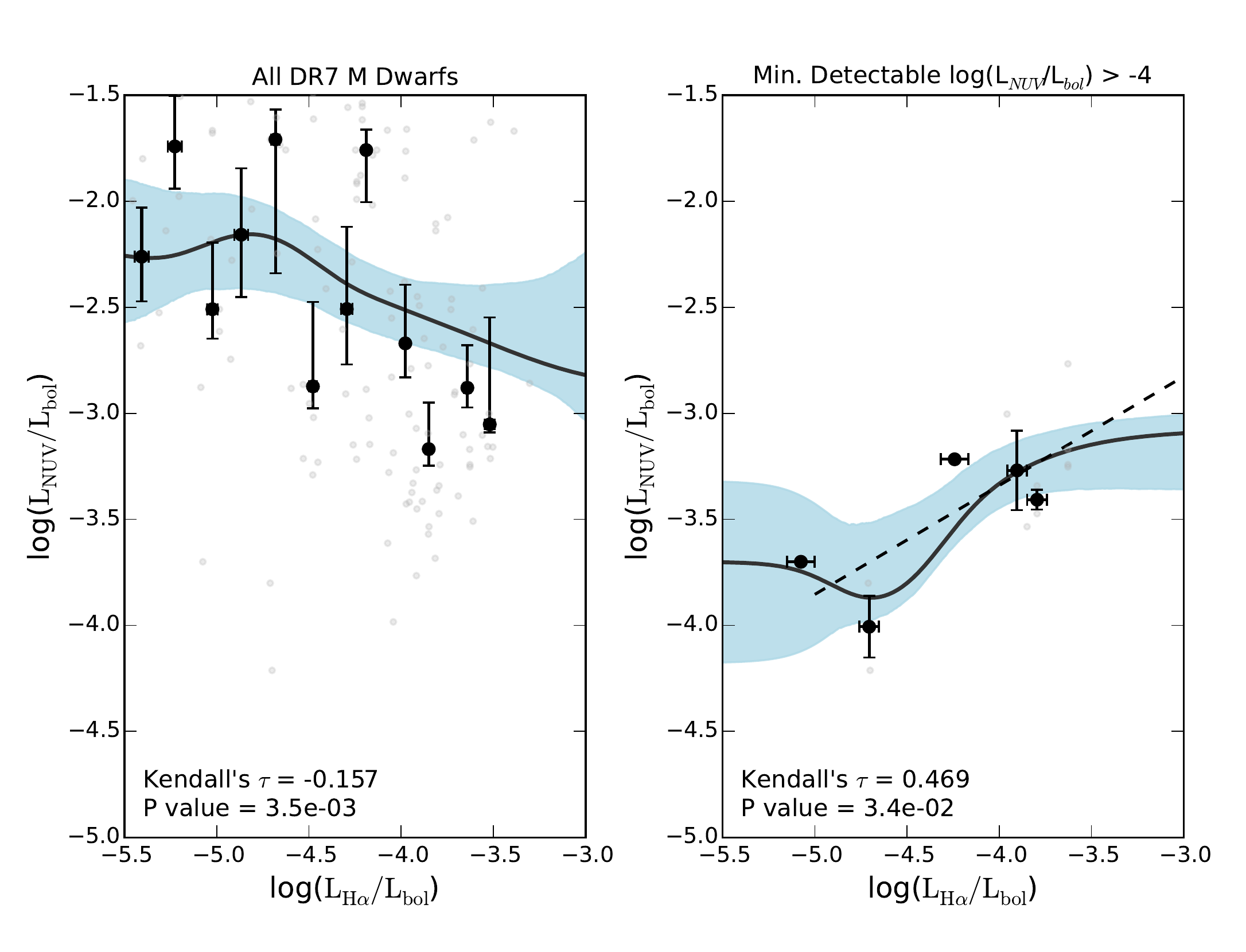}
\caption{The fractional NUV luminosity, log($L_{\mathrm{NUV}}/L_{\mathrm{bol}}$), plotted against the fractional H$\alpha$ luminosity, log($L_{\mathrm{H\alpha}}/L_{\mathrm{bol}}$), for stars in the DR7 M dwarf sample (individual stars and median bins).  There is a low-significance trend for stars with minimum detectable fractional luminosities $>$-4.0 dex (right), with a positive slope at $\sim$1$\sigma$ significance.  The full sample (left) is largely flat, likely due to the highly active stars with possibly ``saturated'' NUV emission levels.  The non-parametric spatial averaging (95\% confidence region in blue with the mean in black) also shows no significant trend.}
\label{fig:DR7 nuvhalpha}
\end{figure}

The smallest DR7 NUV luminosities measured are 1-2 orders of magnitude higher than the smallest PMSU luminosities, due to the smaller distances of the PMSU stars.  Additionally, we observed no correlation between DR7 UV data and H$\alpha$ emission.  This may indicate an upper bound to the amount of quiescent NUV emission that occurs in very active stars, which we will refer to as ``saturation''.  We separated the DR7 sample into groups of similar spectral types to see if our null result was caused by the grouping of spectral types, but found no correlations, even for late spectral types ($\sim$M6$-$M9), which can be observed at smaller distances without saturating the SDSS detectors and may be able to probe lower activity levels.  Figure \ref{fig:DR7 nuvhalpha} shows the null correlation of H$\alpha$ with NUV luminosity for all DR7 spectral types (left). The DR7 data also predict lower H$\alpha$ luminosities for a given NUV luminosity than is seen in the PMSU sample.  Again, this is likely the result of the distance-based selection effect; H$\alpha$ activity is relatively easy to measure throughout the sample, but at large distances only extremely NUV-active stars can be found.

If we restrict our sample to low NUV detection thresholds (Figure \ref{fig:DR7 nuvhalpha}, right panel), there are $\sim$1$\sigma$ hints of a trend, showing consistency with our PMSU results.  However, the saturated M dwarfs at log(L$_{\textrm{NUV}}$/L$_{\textrm{bol}}$) $>$ -3.0 cannot be explained entirely by our 14\% fraction of spurious DR7 matches or a distance bias, and likely result from saturation.  In DR7 it is not possible to draw conclusions with the FUV data due to the small number of stars with measurable FUV flux.

\begin{figure}
\epsscale{1.28}
\plotone{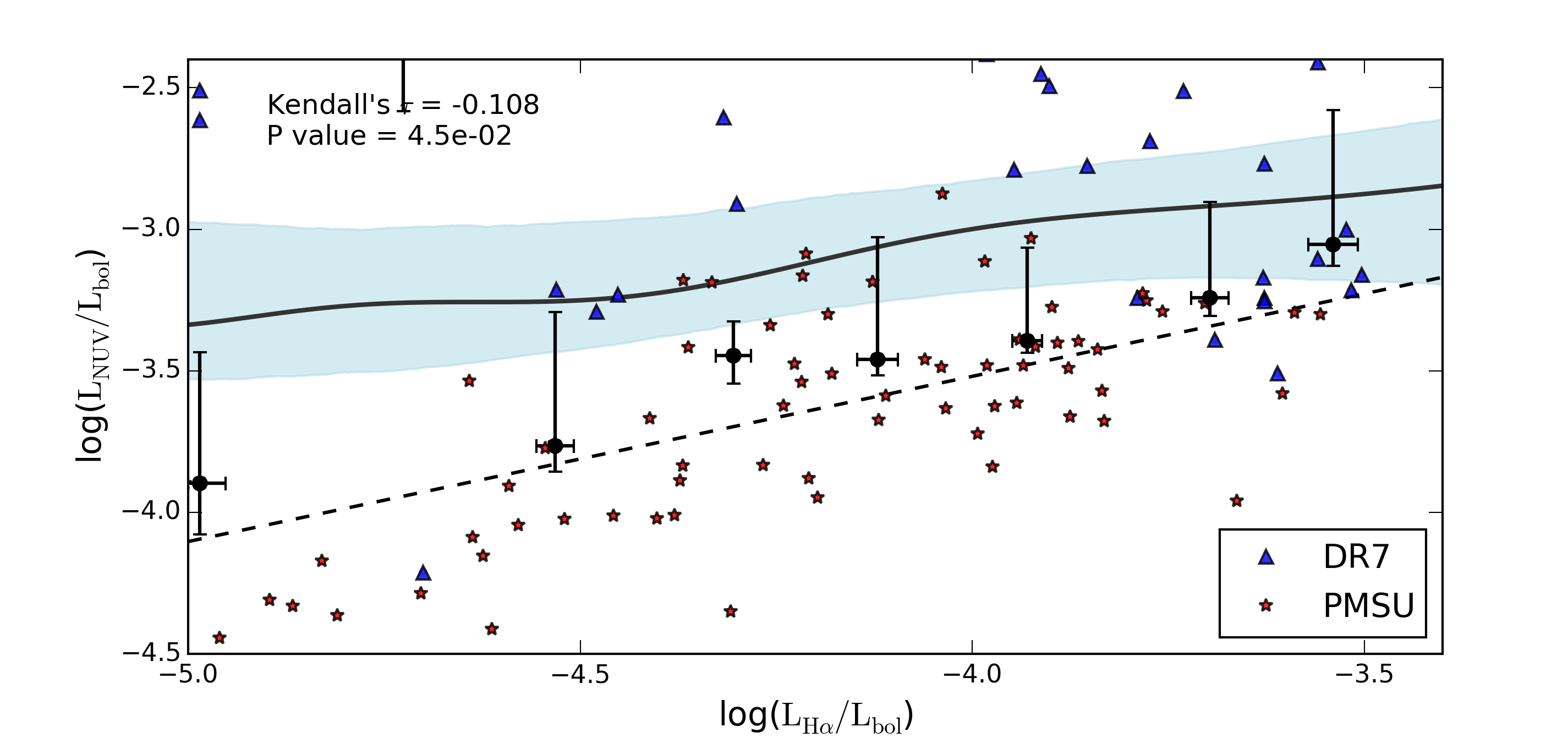}
\caption{The relation between fractional NUV luminosity and fractional H$\alpha$ luminosity for M3 and M4 stars (individual DR7 and PMSU stars and median bins), the most common spectral types, in the combined DR7 and PMSU samples.  At low NUV luminosities only the nearby PMSU sample can be observed, whereas at higher luminosities the highly active DR7 stars are observable.  The best-fit (dashed) line from the PMSU sample is over-plotted.  The non-parametric spatial averaging curve (blue; 95\% confidence with the mean in black) is above the median bins due to the large number of high-luminosity DR7 dwarfs.  Although Kendall's $\tau$ is negative, the median trend appears robust as it is less affected by high-luminosity DR7 dwarfs.}
\label{fig:comb uvhalpha}
\end{figure}

We further demonstrate the high-luminosity behavior of NUV emission by combining the PMSU and DR7 samples for spectral types M3-M4, the most common spectral types in the combined sample, to create a single NUV - H$\alpha$ relation (Figure \ref{fig:comb uvhalpha}).  The low end of NUV activity consists of PMSU stars, while the high end consists mainly of DR7 stars, of which only the most active can be observed.  We over-plotted the best-fit line from PMSU data.  The slope of the NUV-H$\alpha$ relation shown here is consistent with the data, but tends to under-predict the high NUV luminosities, perhaps due to distance selection effects or NUV saturation.  The non-parametric regression is consistently above the median bins due to the effect of highly-active DR7 stars.

\begin{figure*}
\epsscale{1}
\includegraphics[angle=0,width=7in]{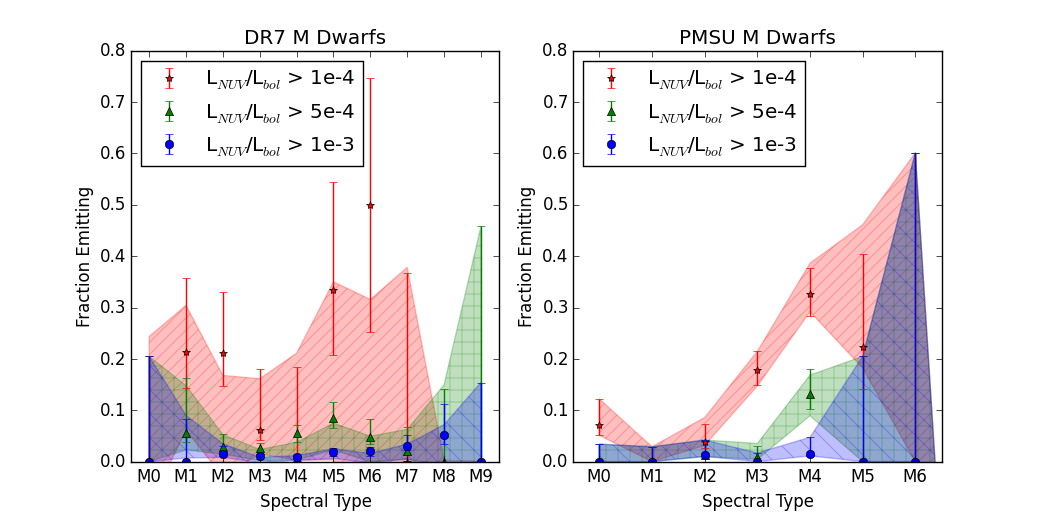}
\caption{The fraction of DR7 and PMSU M dwarfs emitting certain fractions of their luminosity in the NUV as a function of spectral type.  We demonstrate the robustness of this relation by showing the fraction of stars emitting $L_{\mathrm{NUV}}/L_{\mathrm{bol}} > 10^{-4}$ (red stars), $>5\times10^{-4}$ (green triangles), or $10^{-3}$ (blue circles).  Although the emitting fraction is predictably lower for higher NUV minimum luminosities, and the data are correlated across different threshold levels, the fractions universal increase until spectral types $\sim$M4-M6, and the bump at mid-spectral types is consistent with the onset of fully convective stellar interiors in late-type stars.  At later spectral types, the emitting fractions may decrease, but this is a low-significance trend that requires a much larger data set.  The increase to later spectral types could be similar to the H$\alpha$ active fraction, which is a maximum at a spectral type of M8 \citep{W04}.  The maximum NUV-active fraction is $\sim$0.3 for both DR7 and PMSU M dwarfs.  Error bars are computed using binomial statistics, and shaded regions are binomial statistics added in quadrature to uncertainties estimated from Monte Carlo sampling of the $L_{\mathrm{NUV}}/L_{\mathrm{bol}}$ uncertainties.  Our results from Monte Carlo sampling tend to lie lower than the data for DR7, due to the $\sim$18\% distance uncertainty associated with a given emission threshold.}
\label{fig:DR7 lnuv actfrac}
\end{figure*}

\subsection{The Fraction of NUV-Emitting M Dwarfs}
\label{sec:actfrac}

Figure \ref{fig:DR7 lnuv actfrac} shows the fraction of DR7 stars (left) and PMSU stars (right) emitting certain percentages of their luminosity in the NUV as a function of spectral type.  These trends are meant to probe UV magnetic activity, with the caveat that while these measurements are indicative of UV activity, we are unable to directly examine if emission lines were present.  We show three different $L_{\mathrm{NUV}}/L_{\mathrm{bol}}$ thresholds to show that, although the data are somewhat correlated across different threshold levels, the general trends are independent of the specific NUV emission ``cut-off'' levels that we choose.  These values span the range of useful probes of NUV magnetic activity; if we examine stars with $L_{\mathrm{NUV}}/L_{\mathrm{bol}}$ below $10^{-4}$, it becomes clear that we are observing the stellar continuum in early-type dwarfs, as these suddenly have a large NUV-detected fraction while mid- and late-type dwarfs stay the same.  If we examine only stars with $L_{\mathrm{NUV}}/L_{\mathrm{bol}} > 10^{-3}$, there are so few stars with this level of UV emission that we cannot effectively probe the relationship of spectral type to line emission.  Our error bars are computed using binomial statistics.  The shaded regions add binomial statistics in quadrature to errors estimated from Monte Carlo sampling of the  $L_{\mathrm{NUV}}/L_{\mathrm{bol}}$ uncertainties.

Figure \ref{fig:DR7 lnuv actfrac} is noisy, but both data sets show that the fractions of emitting M dwarfs appear to reach a maximum at mid or later spectral types, similar to the trend in H$\alpha$ activity, which is maximized at a spectral type of M8 \citep{W04}.  The emitting fractions show a significant bump between spectral types M3 and M4, where M dwarfs begin to become fully convective \citep{W08}, mirroring the increase in H$\alpha$ activity at these types.  We have excluded PMSU M7-M9 dwarfs due to lack of data, and have little data with which to draw conclusions for M5 and M6 dwarfs.  For early- to mid-spectral types, the NUV-emitting fractions for DR7 stars follow the same general trend as PMSU stars, peaking in M4-M6 dwarfs.  The data are consistent with the H$\alpha$ peak in active fraction, but the uncertainties are too large for late spectral types to determine the most active types.  The lowest threshold shows that $\sim$30-40\% of mid-type M dwarfs have some level of detectable NUV activity.  The true fraction of NUV-active M dwarfs is likely higher, as even our lowest threshold is above the NUV-emitting level of some moderately active M dwarfs such as GJ 876 (Figure \ref{fig:spec}; $L_{\mathrm{NUV}}/L_{\mathrm{bol}} = 1.9\times10^{-5}$) and our strict selection cuts remove some matches.  We examined the FUV-emitting fractions as a function of spectral type, but found that the small samples and resulting error bars were too large to draw any significant conclusions. 

\begin{figure}
\includegraphics[angle=0,width=3.4in]{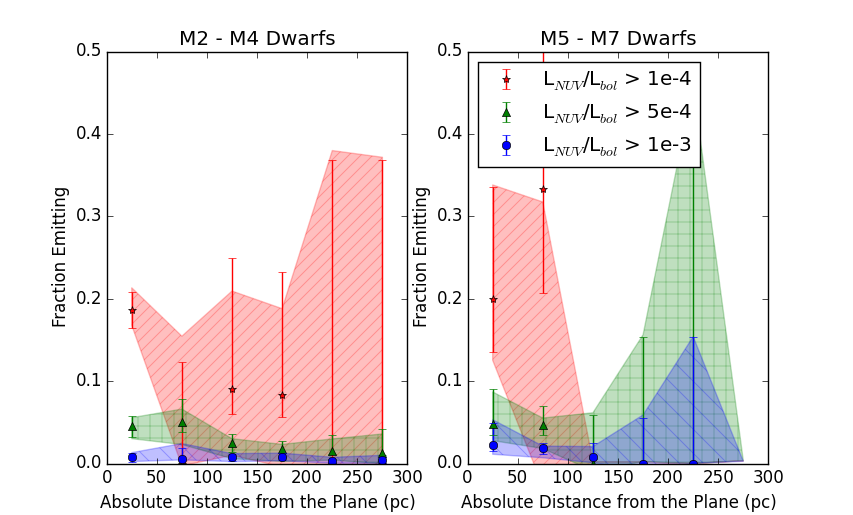}
\caption{The fraction of DR7 M dwarfs emitting certain fractions of their luminosity in the NUV as a function of absolute vertical height above the Galactic plane for two subsets of the most active spectral types (M2 - M4 and M5 - M7).  We show three different NUV emission thresholds: $L_{\mathrm{NUV}}/L_{\mathrm{bol}} > 10^{-4}$ (red circles), $>5\times10^{-4}$ (green stars), or $>10^{-3}$ (blue triangles).  For late-type stars, the data show that the NUV-emitting fractions fall by a factor of 2-4 as distance from the plane increases; for early type stars, we see some evidence for a decrease but our data are noisier likely due to the lower average activity of these stars.  This behavior is roughly consistent with that of H$\alpha$ activity, which is also shown to decrease by a factor of $\sim$2-4 over distances of 0 to 200 pc from the Galactic Plane \citep{W08}.  Too little data are available to investigate how NUV emission in individual or less active spectral types changes with age.}
\label{fig:DR7 lnuv dist}
\end{figure}

We examined the dependence of these NUV-emitting fractions on distance for the most active DR7 spectral types, M3-M7 (Figure \ref{fig:DR7 lnuv dist}), the spectral types for which we have a significant sample.  We combined multiple spectral types to increase our statistics.  Figure \ref{fig:DR7 lnuv dist} shows a decrease in NUV-emitting fraction with absolute vertical distance from the Galactic plane for the higher two thresholds, although the evidence is marginal for early-type stars likely due to their lower active fraction overall.  Due to dynamical heating, stars gravitationally scatter as they move through the Galaxy and thus, older stars are preferentially farther from the plane \citep{W06,W08}.  A typical decrease in H$\alpha$ activity for the same distance from the Galactic plane is a factor of 2-4 \citep{W08}; for M5-M7 dwarfs, the high level of UV activity examined in this figure (green triangles/blue circles) falls off at a similar rate.  This figure also confirms the results of \citet{S11}, who found that UV stellar activity tends to be an indication of youth.

\section{Conclusions}

We investigated the NUV and FUV emission of low-mass stars by identifying 
\drnuvnum\ NUV and \drfuvnum\ FUV matches between GALEX and the DR7 M dwarf sample and \pmsunuvnum\ NUV and \pmsufuvnum\ FUV matches between GALEX and the PMSU M dwarf sample.  We found evidence for clear correlations between optical (H$\alpha$), NUV, and FUV emission strength in the PMSU M dwarf sample.  For PMSU, we found that the relation between the logarithm of the NUV and H$\alpha$ fractional luminosities is well fit by a line of slope $\sim$0.6.  The DR7 sample demonstrates that there may be evidence that highly NUV-active stars reach a ``saturation'' point where increased activity at other wavelengths does not correspond to a large increase in NUV luminosity, but further work is needed to verify this.

By examining the NUV-emitting fractions of these M dwarfs (above three semi-arbitrary thresholds), we found the fractions peak at mid spectral types, with at least $\sim$30-40\% of young M dwarfs having some level of activity.  For the most active spectral types in the DR7 sample (M5-M7), we determined that NUV emission is also a function of stellar age due to the fact that it falls off significantly with increasing distance from the Galactic plane.  This result agrees with previous studies of activity in H$\alpha$ and other wavelength regimes, and is another indication of the effect of stellar age, the strong correlation between activity at various wavelengths across the stellar spectrum, and a confirmation of the conclusion of \citet{S11} that young low-mass stars tend to be active in the ultraviolet.  

Our results indicate that NUV and H$\alpha$ activity in M dwarfs are strongly correlated and exhibit many of the same characteristics.  In the absence of UV spectra for a large sample of M dwarfs, these relations can be used to constrain stellar atmosphere models and the UV radiation incident on extrasolar planets (e.g. \citealp{Rugheimer15}).  Larger UV samples of mid- and late-type M dwarfs are needed to characterize the UV-active fraction and the UV age-activity relation before UV radiation can be characterized as a function of stellar age for these stars.

We are providing our full cross matched catalog to the Astronomical community for future use at {\url{http://www.pha.jhu.edu/~djones/GALEX.html}}.  Future studies will be able to use these relations and data in order to gain a more comprehensive understanding of the stellar physics of low-mass stars and the potential habitability of their attending exoplanets.

\acknowledgments

We would like to thank the anonymous referee for many helpful comments.  We also thank Evgenya Shkolnik, Saurav Dhital, Dylan Morgan, John Bochanski, Suvi Gezari, and Tamas Budavari for their many contributions to this study.  AAW acknowledges the support of the NASA/GALEX grant program under Cooperative Agreement No. NNX10AM62G issued through the NASA Shared Services Center.   A.A.W also acknowledges funding from NSF grants AST-1109273 and AST-
1255568 and the support of the Research Corporation for Science Advancement’s
Cottrell Scholarship.  This study is based on observations made with the NASA Galaxy Evolution Explorer. 
GALEX is operated for NASA by the California Institute of Technology under NASA contract NAS5-98034.

Funding for the Sloan Digital Sky Survey (SDSS) and SDSS-II has been
provided by the Alfred P. Sloan Foundation, the Participating
Institutions, the National Science Foundation, the U.S. Department of
Energy, the National Aeronautics and Space Administration, the
Japanese Monbukagakusho, and the Max Planck Society, and the Higher
Education Funding Council for England. The SDSS Web site is
http://www.sdss.org/.

The SDSS is managed by the Astrophysical Research Consortium (ARC) for
the Participating Institutions. The Participating Institutions are the
American Museum of Natural History, Astrophysical Institute Potsdam,
University of Basel, University of Cambridge, Case Western Reserve
University, The University of Chicago, Drexel University, Fermilab,
the Institute for Advanced Study, the Japan Participation Group, The
Johns Hopkins University, the Joint Institute for Nuclear
Astrophysics, the Kavli Institute for Particle Astrophysics and
Cosmology, the Korean Scientist Group, the Chinese Academy of Sciences
(LAMOST), Los Alamos National Laboratory, the Max-Planck-Institute for
Astronomy (MPIA), the Max-Planck-Institute for Astrophysics (MPA), New
Mexico State University, Ohio State University, University of
Pittsburgh, University of Portsmouth, Princeton University, the United
States Naval Observatory, and the University of Washington.


\clearpage

\begin{turnpage}
\begin{deluxetable}{cccccccccc}
\tablewidth{0pt}
\tablewidth{0pt}
\tablecolumns{9} 
\tabletypesize{\scriptsize}
\tablecaption{DR7 M Dwarfs Found in GALEX\tablenotemark{a}}
\renewcommand{\arraystretch}{1}
\tablehead{
\colhead{SDSS Object ID}&
\colhead{GALEX Object ID}&
\colhead{RA\tablenotemark{b}}&
\colhead{Dec\tablenotemark{b}}&
\colhead{NUV Mag}&
\colhead{L$_{\textrm{NUV}}$/L$_{\textrm{bol}}$}&
\colhead{FUV Mag}&
\colhead{L$_{\textrm{FUV}}$/L$_{\textrm{bol}}$}&
\colhead{L$_{H\alpha}$/L$_{\textrm{bol}}$}&
\colhead{Sp. Type}}
\startdata
587728950122053760&2411926227589794816&10:06:52.421&+01:02:28.07&22.04$\pm$0.19&1.52e-02$\pm$2.60e-03&22.36$\pm$0.24&8.81e-03$\pm$1.97e-03&3.05e-05$\pm$2.52e-05&M1\\
587724648192409856&2415409480426589184&13:23:09.313&-03:23:39.34&22.23$\pm$0.21&1.74e-02$\pm$3.32e-03&22.44$\pm$0.24&1.13e-02$\pm$2.46e-03&5.36e-05$\pm$2.11e-05&M6\\
587725489987321984&2415866877263743488&16:56:51.488&+60:55:32.75&23.00$\pm$0.27&3.93e-03$\pm$9.74e-04&23.01$\pm$0.28&3.02e-03$\pm$7.70e-04&-1.21e-05$\pm$8.42e-06&M2\\
587725578034413696&2415972430377917440&17:25:52.453&+63:29:06.62&21.73$\pm$0.10&1.89e-02$\pm$1.78e-03&\nodata&\nodata&-7.73e-06$\pm$7.15e-06&M2\\
587731185135976704&2417450174007744512&00:53:19.471&-00:51:43.71&23.19$\pm$0.30&7.17e-03$\pm$1.96e-03&23.52$\pm$0.42&4.11e-03$\pm$1.58e-03&-4.17e-05$\pm$1.92e-05&M3\\
587724198819921920&2419174208237998080&01:50:16.365&+14:05:28.68&23.17$\pm$0.31&2.11e-04$\pm$6.02e-05&\nodata&\nodata&1.91e-04$\pm$6.11e-05&M2\\
588009368547033216&2421496376795863040&09:11:55.458&+54:01:26.79&20.45$\pm$0.04&3.11e-02$\pm$1.28e-03&\nodata&\nodata&-4.64e-05$\pm$1.33e-05&M0\\
587724242309415168&2423853729725811712&04:08:57.875&-04:41:09.30&23.02$\pm$0.26&1.38e-02$\pm$3.32e-03&\nodata&\nodata&-1.68e-04$\pm$8.31e-05&M3\\
588007005258907776&2431594291587456000&14:54:44.247&+59:27:53.81&22.45$\pm$0.23&3.77e-03$\pm$7.99e-04&23.59$\pm$0.26&1.02e-03$\pm$2.42e-04&1.56e-05$\pm$4.77e-06&M7\\
587729228228067456&2468819357257443328&16:22:09.325&+50:07:52.51&19.10$\pm$0.02&5.06e-02$\pm$8.87e-04&18.72$\pm$0.03&5.55e-02$\pm$1.32e-03&1.53e-04$\pm$3.75e-05&M0\\
\enddata
\tablenotetext{a}{The full GALEX - DR7 matched source catalog can be found online at {\url{http://www.pha.jhu.edu/~djones/GALEX.html}}.}
\tablenotetext{b}{The positions reported here are taken from the SDSS DR7, which contains more precise astrometry than GALEX.}
\label{table:DR7 data}
\end{deluxetable}

\begin{deluxetable}{cccccccccc}
\tablewidth{0pt}
\tablewidth{0pt}
\tablecolumns{9} 
\tabletypesize{\scriptsize}
\tablecaption{PMSU M Dwarfs Found in GALEX\tablenotemark{a}}
\renewcommand{\arraystretch}{1}
\tablehead{
\colhead{SDSS Object ID}&
\colhead{GALEX Object ID}&
\colhead{RA\tablenotemark{b}}&
\colhead{Dec\tablenotemark{b}}&
\colhead{NUV Mag}&
\colhead{L$_{\textrm{NUV}}$/L$_{\textrm{bol}}$}&
\colhead{FUV Mag}&
\colhead{L$_{\textrm{FUV}}$/L$_{\textrm{bol}}$}&
\colhead{L$_{H\alpha}$/L$_{\textrm{bol}}$}&
\colhead{Sp. Type}}
\startdata
Gl 2&6372076894527948800&00:05:10.888&+45:47:11.64&18.73$\pm$0.06&4.70e-05$\pm$1.41e+00&\nodata&\nodata&-2.66e-06$\pm$9.61e+05&M1\\
V351&6375982386675452928&00:08:27.279&+17:25:27.47&19.31$\pm$0.06&4.54e-05$\pm$5.66e+00&21.56$\pm$0.29&4.44e-06$\pm$5.66e+00&1.66e-06$\pm$1.74e+06&M0\\
G131-026&6375982351239873536&00:08:53.919&+20:50:25.24&18.57$\pm$0.07&6.57e-04$\pm$4.24e+01&\nodata&\nodata&7.46e-05$\pm$3.60e+08&M4\\
G131-047&6375982355536939008&00:16:56.779&+20:03:55.10&21.00$\pm$0.18&1.37e-04$\pm$4.24e+01&22.74$\pm$0.47&2.14e-05$\pm$4.24e+01&9.90e-06$\pm$1.82e+08&M3\\
LHS1054&6376721218758771712&00:17:20.323&+29:10:58.81&20.64$\pm$0.25&7.24e-05$\pm$8.48e+00&\nodata&\nodata&2.67e-06$\pm$1.23e+07&M2\\
G158-052&6380310079474761728&00:17:40.903&-08:40:56.15&19.22$\pm$0.06&1.27e-04$\pm$1.13e+01&21.52$\pm$0.27&1.20e-05$\pm$1.13e+01&7.66e-07$\pm$3.49e+06&M0\\
Gl 16&6380978543887648768&00:18:16.589&+10:12:10.03&20.16$\pm$0.15&3.65e-05$\pm$4.24e+00&\nodata&\nodata&-2.14e-06$\pm$4.23e+06&M1\\
G130-063&6376721273519606784&00:18:53.530&+27:48:50.00&21.27$\pm$0.21&1.31e-04$\pm$4.24e+01&\nodata&\nodata&-1.31e-05$\pm$2.57e+08&M4\\
GJ 2003&6380415634736415744&00:20:08.380&-17:03:40.97&20.98$\pm$0.18&1.96e-05$\pm$9.90e+00&\nodata&\nodata&-3.24e-06$\pm$4.55e+06&M0\\
LP149-56&6372076865539016704&00:21:57.81&+49:12:38.00&19.34$\pm$0.08&3.48e-04$\pm$4.10e+01&20.91$\pm$0.21&6.39e-05$\pm$4.10e+01&5.34e-05$\pm$8.61e+07&M2\\
\enddata
\tablenotetext{a}{The full GALEX - PMSU matched source catalog can be found online at {\url{http://www.pha.jhu.edu/~djones/GALEX.html}}.}
\tablenotetext{b}{The positions reported here are taken from the SIMBAD Astronomical Database.}
\label{table:PMSU data}
\end{deluxetable}
\end{turnpage}

\clearpage

\end{document}